# Dielectrophoretic assembly of liquid-phase-exfoliated $TiS_3$ nanoribbons for photodetecting applications

R. Frisenda*[1], E. Giovanelli[1], P. Mishra[1], P. Gant[1], E. Flores[2], C. Sánchez[2], J. R. Ares[2], D. Perez de Lara[1], I. J. Ferrer[2], E. M. Pérez*[1], A. Castellanos-Gomez*[3]

[1] Instituto Madrileño de Estudios Avanzados en Nanociencia (IMDEA-nanociencia), Campus de Cantoblanco, E-28049 Madrid, Spain.

[2] Materials of Interest in Renewable Energies Group (MIRE Group), Dpto. de Física de Materiales, Universidad Autónoma de Madrid, E-28049 Madrid, Spain.

[3] Instituto de Ciencia de los Materiales de Madrid (ICMM-CSIC), E-28049 Madrid, Spain.

* E-mail: riccardo.frisenda@imdea.org; emilio.perez@imdea.org; andres.castellanos@csic.es.

**ABSTRACT**

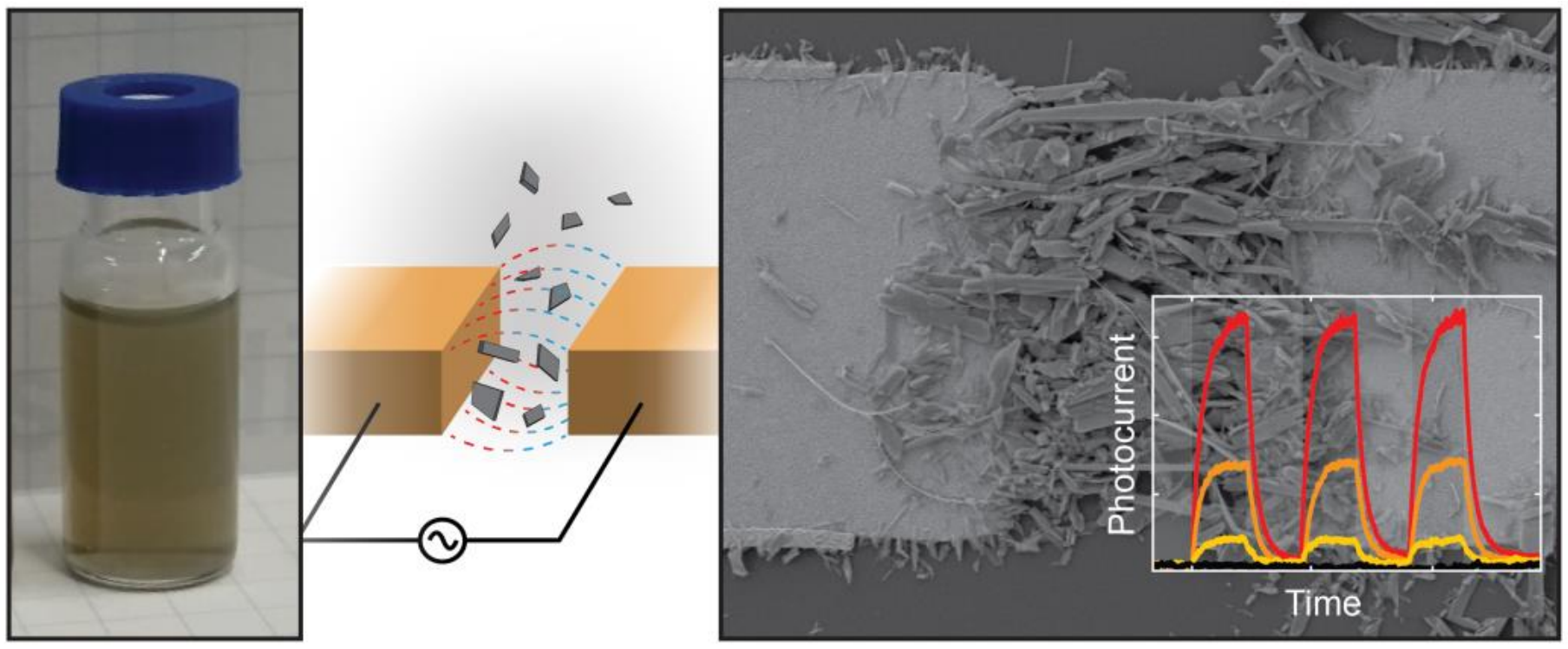

Liquid-phase-exfoliation is a technique capable of producing large quantities of two-dimensional material in suspension. Despite many efforts in the optimization of the exfoliation process itself not much has been done towards the integration of liquid-phase-exfoliated materials in working solid-state devices. In this article, we use dielectrophoresis to direct the assembly of liquid-phase-exfoliated TiS3 nanoribbons between two gold electrodes to produce photodetectors working in the visible. Through electrical and optical measurements we characterize the responsivity of the device and we find values as large as 3.8 mA/W, which improve of more than one order of magnitude on the state-of-the-art for devices based on liquid-phase-exfoliated two-dimensional materials assembled by drop-casting or ink-jet methods.

## MAIN TEXT

Right after the first works on mechanically exfoliated two-dimensional (2D) materials[1-3], other preparation methods were rapidly explored and developed, oriented toward larger scale production. A significant example of a bottom-up approach is chemical vapor deposition (CVD) that provides a way to synthesize high quantities of homogenous materials with large crystalline domains[4-6], but is costly and complicate. From the top-down perspective, liquid-phase-exfoliation (LPE) is an appealing route to produce large amounts of 2D materials in the form of small crystallites suspended in a solvent[7-13]. LPE represents a cheap technique that can be easily scaled up and with a strong potential to be used in printed electronics.

Significant efforts have been made regarding the optimization of the LPE process[14-19], mostly aimed at controlling the thickness and lateral size of the flakes and increasing the concentration of the suspensions. However, the integration of LPE materials into functional devices still remains largely unexplored, with the exception of devices based on graphene inks[20, 21]. Additionally, Withers *et al.*[22] showed that heterostructures could be assembled from liquid-phase-exfoliated 2D crystals and they designed a photodetector with a gate-tunable photoresponsivity of up to 43 µA/W. In a different study, Yang *et al.*[23] incorporated liquid-phase-exfoliated materials as dielectric in graphene-based photodetectors. Zhu *et al.*[24] reported similar findings on the outstanding dielectric properties of liquid inks made from thickness-sorted 2D hexagonal boron nitride.

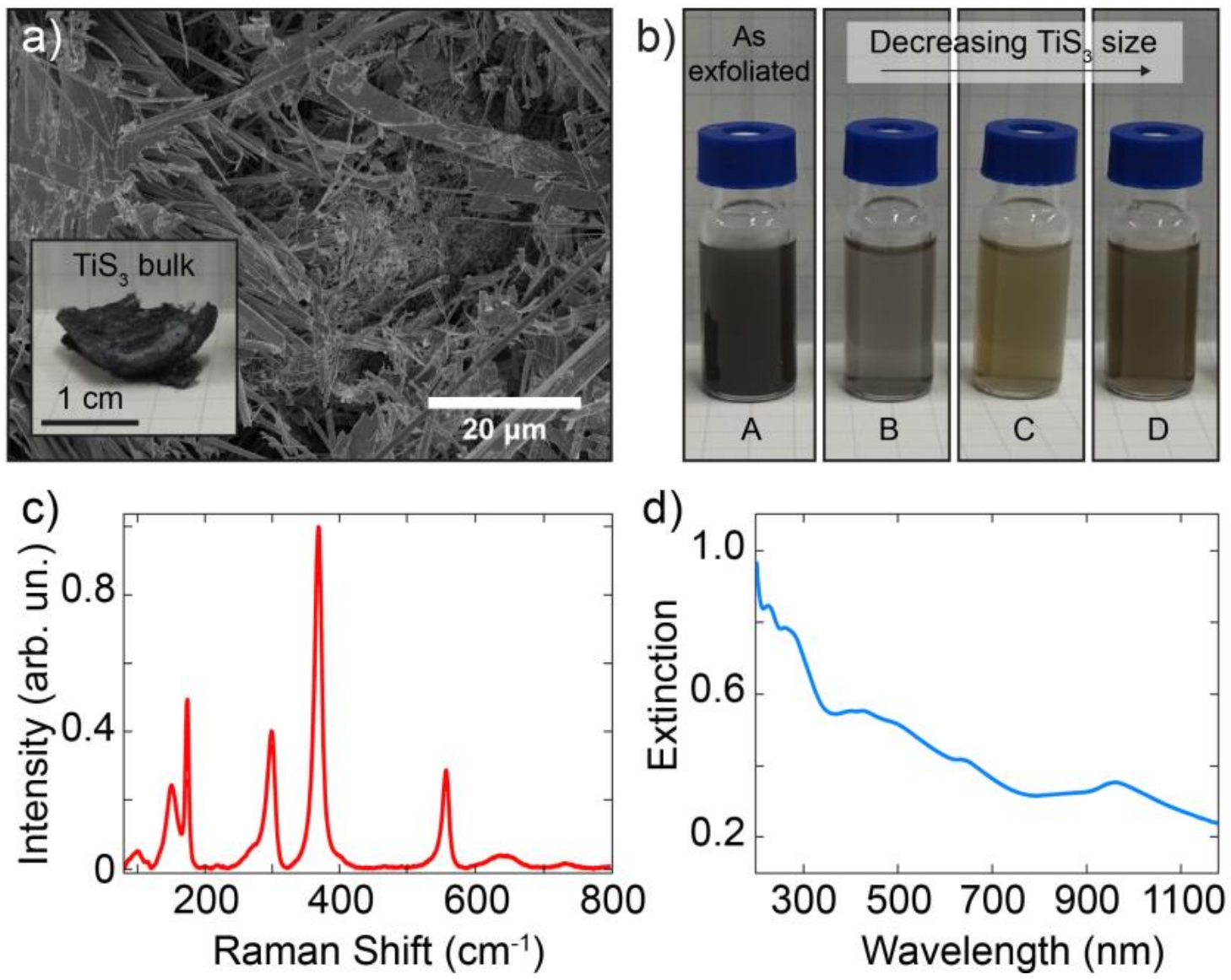

**Figure 1:** a) SEM image of the lamellar $TiS_3$ bulk material. Inset: photograph of bulk $TiS_3$. b) Optical image colloidal suspensions of exfoliated $TiS_3$ in IPA. c) Raman spectrum (normalized at 368 $cm^{-1}$) and d) UV-Vis-NIR spectrum of colloidal suspension D.

Here, we demonstrate how titanium trisulfide ($TiS_3$) nanoribbons obtained by LPE can be assembled between two metallic electrodes by dielectrophoresis (DEP) to fabricate photodetectors. $TiS_3$ is a layered material belonging to the family of semiconducting transition metal trichalcogenides and exhibiting pronounced in-plane anisotropy[25] and great potential to be used in optoelectronics[26-28]. In the present paper we use DEP-assisted deposition to assemble liquid-phase-exfoliated $TiS_3$ nanoribbons in between metallic electrodes to produce sol-id-state photodetectors. The device based on colloidal $TiS_3$ shows responsivities as large as 3.8 mA/W improving on state-of-the-art LPE-based devices. The novel deposition method based on DEP, together with the excellent photodetecting properties of $TiS_3$ retained in the liquid-phase-exfoliated material, paves the route for novel and cheap photodetectors which can be easily scaled up for industrial applications.

Bulk $TiS_3$ was prepared by surface sulfurization of bulk Ti powder at 500 °C, similarly to the procedure described in Ref.[29]. Figure 1a shows a scanning electron microscopy (SEM) picture of lamellar bulk $TiS_3$ and in the inset a photograph of as-synthetized bulk $TiS_3$ (Fig. S1, ESI). We prepared the $TiS_3$ suspensions by ultrasonication of the bulk material in isopropyl alcohol (IPA). We employed successive centrifugation-redispersion sequences to sort out the suspensions according to $TiS_3$ crystallite dimensions. Figure 1b shows a photograph of four suspensions of $TiS_3$ in IPA. From suspension A, resulting directly from the sonication step, were prepared three other suspensions labelled B, C and D.

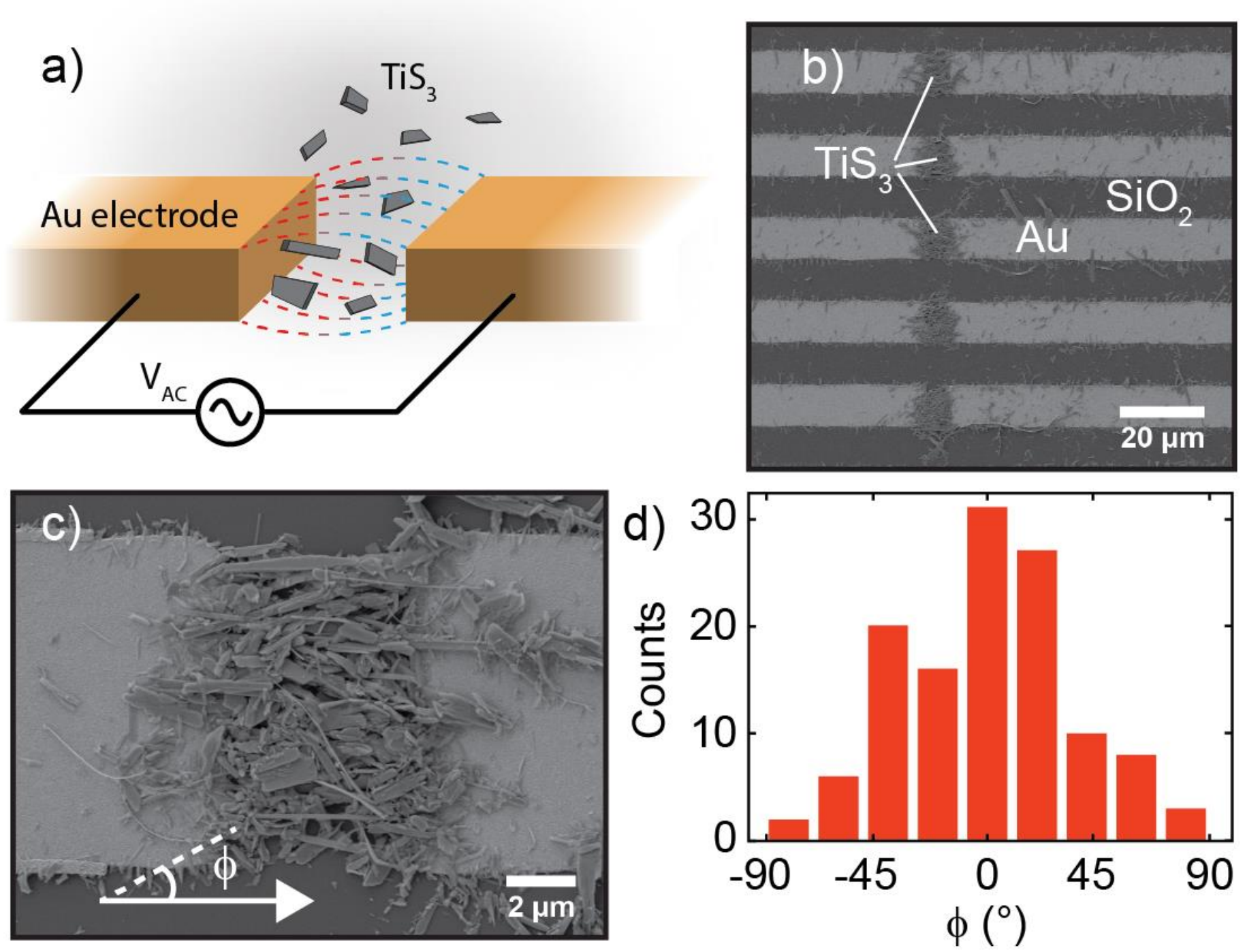

**Figure 2:** a) Sketch of the DEP technique used to fabricate $TiS_3$-based photodetecting devices from the liquid-phase-exfoliated material. b-c) SEM images of aligned $TiS_3$ nanowires between gold electrodes assembled from suspension D. d) Angular distribution of the assembled $TiS_3$ nanoribbons where 0° correspond to a nanoribbon oriented along the source-drain electrodes direction. The angle was measured from 123 nanoribbons identified in the SEM image of panel (c).

The centrifugation procedure led to the following order of decreasing particle weight in the suspension: B > C > D, whereas suspension A contains all possible particle dimensions. The different suspensions are stable from a few hours to several weeks and any of the suspensions can be easily redispersed by simple stirring, which enables long-term use of the colloids.

Figure 1c shows the Raman spectrum of suspension D measured with an excitation wavelength of 532 nm and an optical power of 2 mW. Raman spectra of all four colloids present similar features (Fig. S2-S4, ESI) indicating that LPE results in analogous structures. The Raman spectra are in agreement with those of bulk[30-33] and with those of mechanically exfoliated few-layer $TiS_3$[34], with additional bands that indicate the presence of $TiO_2$ in the form of the anatase polytype[35, 36], probably due to surface/edge oxidation of $TiS_3$. Raman analysis of the bulk confirms that $TiO_2$ is to be found right from the beginning in the starting material, but further oxidation during the sonication process cannot be excluded[37].

The optical properties of liquid-phase-exfoliated $TiS_3$ were studied by UV-Vis-NIR spectroscopy (Fig. 1d and Fig. S5, ESI). These absorption properties are in agreement with the optical bandgap of 1.07 eV (corresponding to an absorption onset of 1160 nm) reported for bulk as synthesized material[38]. This is expected as the band structure of $TiS_3$ is mostly thickness independent[38, 39]. The Tauc plot (Fig. S6, ESI), which is generated from the absorption data of suspension D, confirms that the $TiS_3$ nanoribbons have a direct optical bandgap of 1.02 eV, consistent with values from the literature[29, 38, 40].

In order to integrate liquid-phase-exfoliated $TiS_3$ into functional devices, we employed a DEP-assisted deposition. In the DEP phenomenon, a non-uniform electric field polarizes the suspended particles of a colloid within the corresponding solvent and gradients in the electric field force the alignment of the particles (dipoles) with the local field lines and push these particles towards regions of most intense field. The relative polarizabilities of both components of the colloid (suspended particles and solvent) are crucial for this process. In previously studies, DEP has been used successfully as a method to assemble carbon nanotubes[41] or graphene[42-44] between metallic electrodes, but its use with other two-dimensional materials is rather scarce and very recent[28].

In our implementation of the technique, sketched in Fig. 2a, a 1 MHz sinusoidal signal of 10 V peak-to-peak is applied between drain and source electrodes, which are separated by 5 µm. A drop of the $TiS_3$ colloidal suspension is casted onto the electrodes and the electric field generated by the oscillating source-drain voltage polarizes the nanoribbons with respect to the solvent. Gradients in the electric fields push these nanoribbons toward the region of highest electric field. $TiS_3$ nanoribbons are thus deposited in a controlled way, preferentially bridging the electrodes, between which the electric field intensity is the highest. This allows the fabrication of a fully packed $TiS_3$ channel out of a suspension of relatively low concentration. $TiS_3$ LPE was performed in IPA due to its ability to efficiently stabilize the resulting colloidal nanoribbons, its relatively low boiling point, and its weak polarizability (≈7 Å$^3$, see Ref.[45]). The DEP force acting on tube-shaped particle (which is a good approximation to the $TiS_3$ nanoribbons) can be written as:

$$F_{DEP} \propto \varepsilon_m \, \mathrm{Re}[(\varepsilon_p^* - \varepsilon_m^*)/(\varepsilon_m^*)] \, \nabla|E|^2$$

where $\varepsilon_p^*$ and $\varepsilon_m^*$ are the complex permittivity of the particles and the suspension medium respectively, and E is the electric field. If a suspended particle has polarizability higher than the medium, $|\varepsilon_p^*| > |\varepsilon_m^*|$, the force will push the particle in the regions where the electric field intensity is the highest.

From now on in the main text we will focus on the DEP-based assembly from suspension D, which contains the smallest particles. A SEM image of the resulting device is reproduced in Figure 2b, and Figure 2c shows a zoom in the region separating one particular pair of electrodes. From these images, it can be seen that most of the $TiS_3$ nanoribbons have accumulated in the region between the electrodes, where the source and drain electrodes are facing each other, with only a small amount of material deposited on the side of the electrodes or on the $SiO_2$ surface. Moreover, thanks to DEP-assisted deposition, $TiS_3$ nanoribbons mostly adopt an orientation that is parallel to the source-drain direction as confirmed by the analysis of the angular distribution of the nanoribbons in relation to the electrodes shown in Fig. 2d, where the peak at 0° indicates that the preferred $TiS_3$ configurations are those in which the nanoribbon length is parallel to the source-drain direction (Fig. S7, ESI). A distribution of the nanoribbons length is shown in Fig. S8 of the ESI.

After the DEP assembly, we characterized the electrical transport of the fabricated device at room temperature and under high-vacuum conditions ($P \approx 2 \cdot 10^{-6}$ mbar). Figure 3a displays the current-voltage (*I-V*) characteristics of the device recorded in the dark and under illumination with a blue laser (λ = 405 nm). The laser has been focused on the device surface in a spot size of approximately 120 µm of diameter. In dark conditions, the *I-V* is asymmetric in bias and the maximum current flowing through the device reaches approximately 25 µA at a bias voltage of 1 V. Upon illumination, the device shows an enhancement of the current because of the photogeneration of charge carriers, visible in the I-Vs of Figure 3a recorded for increasing laser intensities. For samples with a large amount of loosely connected nanoribbons the electrical characteristics turned out not to be reproducible because of continuous switching that we attribute to electromechanical motion of the nanoribbons. See for example a device built from suspension A (Fig. S9-S10, ESI) which shows frequent switches in current with and without illumination.

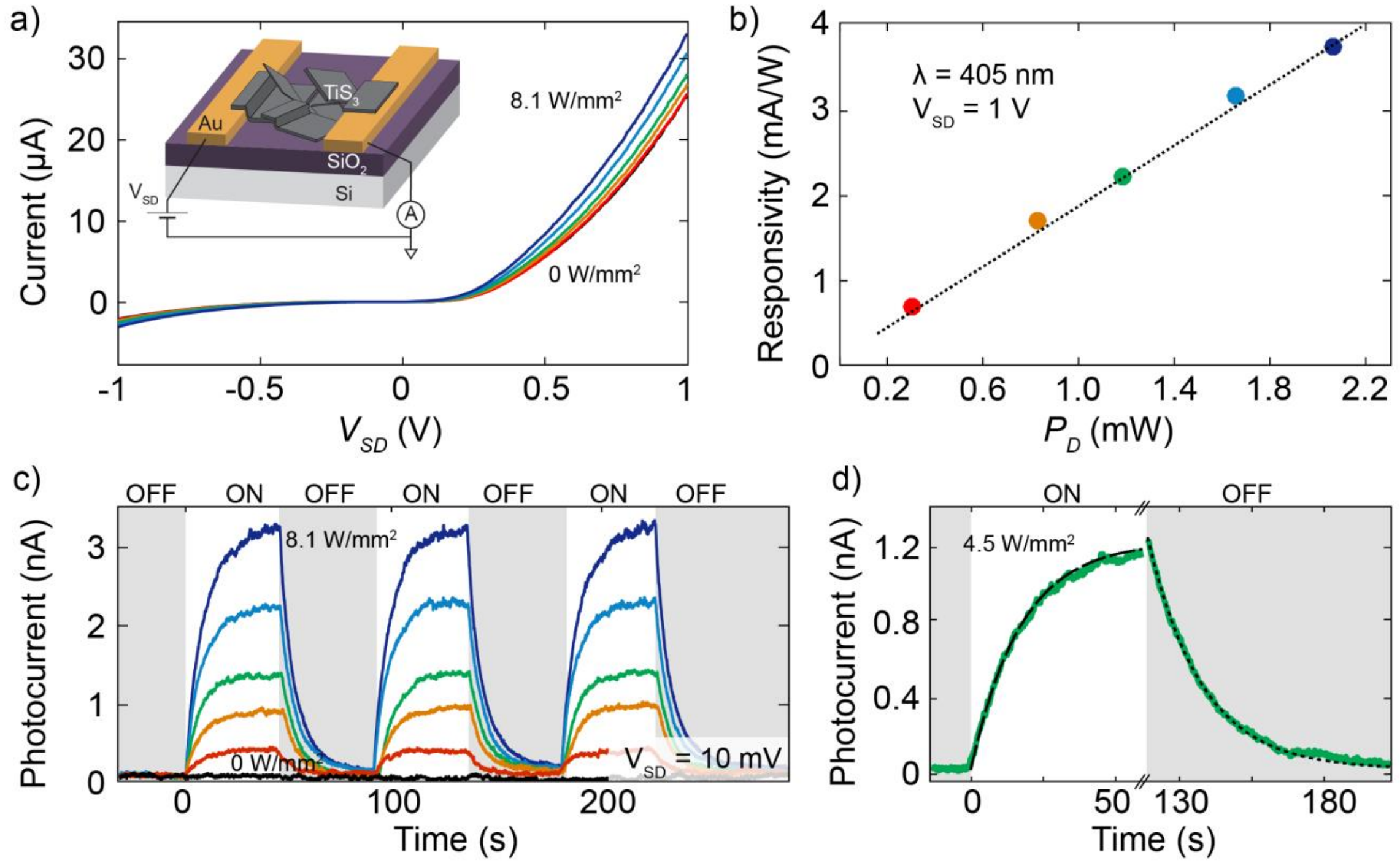

**Figure 3:** a) Current-voltage characteristics of the LPE TiS3 device in dark and upon illumination with a laser (wavelength 405 nm) for increasing light power densities up to 8.1 W/mm$^2$. Inset: schematic illustration of the fabricated device. b) Responsivity at a bias voltage of 1 V as a function of light power. The dashed line is a linear fit to the data. c) Time response of the photocurrent recorded at a bias voltage of 10 mV upon a modulated optical excitation (frequency 8 mHz) for increasing laser power densities up to 8.1 W/mm$^2$. d) Time response at 4.5

W/mm$^2$ of optical power density. The dashed lines are fit to an exponential decay function; the time constant of the system is 8 ± 2 s.

The generated photocurrent in the device made form suspension D reaches a value of 7.4 µA at 1 V with a light/dark current ratio of 1.3. Figure 3b displays the responsivity of the device at a bias of 1 V as a function of the incident laser power $P_D$. The responsivity at each laser power is calculated from the *I-V*s dividing the photocurrent at a voltage of 1 V by the product of the total incident illumination power density and the active area of the device. The largest responsivity is 3.8 mA/W, recorded for the largest excitation power $P_D$. The extracted responsivity is low when compared to that of a single $TiS_3$ nanoribbon[26] which is many order of magnitude higher. This can be explained by the fact that in the present case the transport in the device is dominated by the $TiS_3$/$TiS_3$ interfaces and not by the intrinsic $TiS_3$ material properties giving lower overall performances, as was already observed[18] in the case of LPE $MoS_2$. Interestingly, the responsivity shows a pronounced linear dependence on the excitation power. This monotonous increase in responsivity for increasing incident power is relatively unusual in low-dimensional semiconducting photodetectors, for which a monotonous decrease as a function of power is usually observed. Such a linear increasing behavior has been explained theoretically in other semiconducting devices by the presence of recombination centers with different intragap energies in the semiconductor[18, 46]. The incident light shifts the Fermi level thus changing the occupancy of these centers, which can result in a larger lifetime of the carriers at higher incident power densities. Due to the LPE method our devices are likely to contain various defects, which could account for the linear dependence of the responsivity on the excitation power.

Finally, in order to study the time response of the device we measured the source-drain current as a function of time while switching on and off the illumination. Figure 3c shows the photocurrent recorded during several switching cycles at 10 mV of bias voltage and with increasing illumination power. The response time extracted from Fig. 3d is τ = (8 ± 2) s. In addition, we studied the photoresponse with a laser of 638 nm in order to rule out a possible contribution to the photoresponse due to the presence of $TiO_2$ given its negligible response for excitation wavelengths longer than 500 nm[47]). The small variation in the responsivity at 638 nm and 405 nm (Fig. S11, ESI) is in agreement with the energy dependence of $TiS_3$ nanoribbons absorptivity excluding significant contributions of $TiO_2$ to the photoresponse of the device.

In conclusion, we studied a solid-state photodetector based on liquid-phase-exfoliated $TiS_3$ nanoribbons. We showed that a dielectrophoresis-based deposition method can produce high-quality devices assembling the $TiS_3$ ink onto metallic electrodes. Electrical and optical measurements reveal that the responsivities and the general performances of our devices improve on the state-of-the-art of photodetectors based on liquid-phase-exfoliated materials.

**ACKNOWLEDGEMENTS**

We acknowledge funding from the European Commission (Graphene Flagship, CNECTICT-604391). A.C.G. acknowledges financial support from the MINECO (Ramón y Cajal 2014 RYC-2014-01406 and MAT2014-58399-JIN) and from the Comunidad de Madrid (MAD2D-CM Program (S2013/MIT-3007)). R.F. acknowledges support from NWO through the research program Rubicon with project number 680-50-1515. D.PdL. acknowledges the support of MICINN/MINECO (Spain) through the program FIS2015-67367-C2-1-P. E.M.P. acknowledges funding from the European Research Council MINT (ERC-StG-2012-307609), the MINECO of Spain (CTQ2014-60541-P) and the Comunidad de Madrid (MAD2D project, S2013/MIT-3007). E.G. acknowledges support from the AMAROUT II program (Marie Curie Action, FP7-PEOPLE-2011-COFUND (291803)).

# Supporting Information: Dielectrophoretic assembly of liquid-phase-exfoliated $TiS_3$ nanoribbons for photodetecting applications

R. Frisenda*[1], E. Giovanelli[1], P. Mishra[1], P. Gant[1], E. Flores[2], C. Sánchez[2], J. R. Ares[2], D. Perez de Lara[1], I. J. Ferrer[2], E. M. Pérez*[1], A. Castellanos-Gomez*[3]

[1] Instituto Madrileño de Estudios Avanzados en Nanociencia (IMDEA-nanociencia), Campus de Cantoblanco, E-28049 Madrid, Spain.

[2] Materials of Interest in Renewable Energies Group (MIRE Group), Dpto. de Física de Materiales, Universidad Autónoma de Madrid, E-28049 Madrid, Spain.

[3] Instituto de Ciencia de los Materiales de Madrid (ICMM-CSIC), E-28049 Madrid, Spain.

* E-mail: riccardo.frisenda@imdea.org; emilio.perez@imdea.org; andres.castellanos@csic.es.

**MATERIALS AND METHODS**

*Preparation of colloidal suspensions by liquid-phase exfoliation.* 10 mL of isopropanol (IPA) were added to 10 mg of bulk $TiS_3$, and the heterogeneous mixture was stirred and sonicated for 1 min. The obtained dispersion was sonicated for 1 h in a Fisher Scientific FB 15051 ultrasonic bath (37 kHz, 280 W, ultrasonic peak max. 320 W, sine-wave modulation) thermostated at 20 °C. 750 µL of the resulting suspension (suspension A) were kept for further experiments and characterization. The rest of the suspension was centrifuged (990 *g*, 30 min, 25 °C, Beckman Coulter Allegra® X-15R, FX6100 rotor, radius 9.8 cm). The resulting sediment was redispersed (30 s sonication) in 10 mL *i*PrOH (suspension B), while the supernatant (including partially dispersed light sediment) was subjected to a second round of centrifugation (990 *g*, 30 min, 25 °C). The supernatant was discarded and the sediment was redispersed in 1 mL *i*PrOH (suspension D). 750 µL of suspension B were kept for further studies and the rest of the suspension was centrifuged at lower speed (30 *g*, 30 min, 25 °C). ~8 mL of the supernatant were carefully isolated (suspension C). The remaining supernatant and sediment were discarded.

*Scanning Electron Microscopy.* Samples and devices were observed with a Auriga FE-SEM microscope operated at 3 kV in Secondary Electron (SE) mode (in-lens for Figure 1a of the main text; conventional Everhart-Thornley for all other figures).

*Raman spectroscopy.* Samples were analyzed with a Bruker Senterra confocal Raman microscopy instrument (Bruker Optik, Ettlingen, Germany) under the following conditions: objective NA 0.75, 50×; laser excitation: 532 nm, 2 mW. As-synthesized bulk $TiS_3$ was pressed on a glass slide and the data shown correspond to typical spec-tra acquired in different regions of the solid sample. Colloidal suspensions were dried on glass slides at 40 °C and their respective spectra result from the average of 10 measurements performed over their surface.

*UV-Vis-NIR spectroscopy.* Colloidal suspensions were diluted in IPA and their extinction spectrum (both absorption and scattering are measured using such a piece of equipment) was recorded with a Cary 5000 spectrophotometer, Agilent Technologies (wavelength range: 175-3300 nm).

*Device fabrication.* The parallel interdigitated electrodes were fabricated on a silicon (Si) wafer covered with a 300 nm CVD-grown silicon oxide ($SiO_2$) layer. The interdigitated pattern was developed on the Si/$SiO_2$ substrate via a standard photolithography process. A 10 nm chromium layer (adhesion material) and a 100 nm gold layer (electrode material) were deposited onto the developed surface by thermal evaporation. Lift-off was carried out for the gold layer through wet etching of the unexposed surface. The channel gap between the parallel electrodes is 5 µm.

*Dielectrophoresis.* The DEP process was conducted over the interdigitated electrodes structures. The set-up consists in a function generator used to create an oscillating electric field between the electrodes and in an oscilloscope used to monitor the voltage change in real time during the DEP process. The frequency and the peak-to-peak voltage of the AC signal are 1 MHz and 10 V, respectively. As for the deposition and horizontal alignment of $TiS_3$, a droplet of suspended $TiS_3$ nanoribbons was placed onto the electrode gaps in the presence of the non-uniform electric field. The applied electric field polarizes the nanoribbons, thus creating dipoles that align along the electric field lines.

**SEM ANALYSIS OF BULK $TiS_3$**

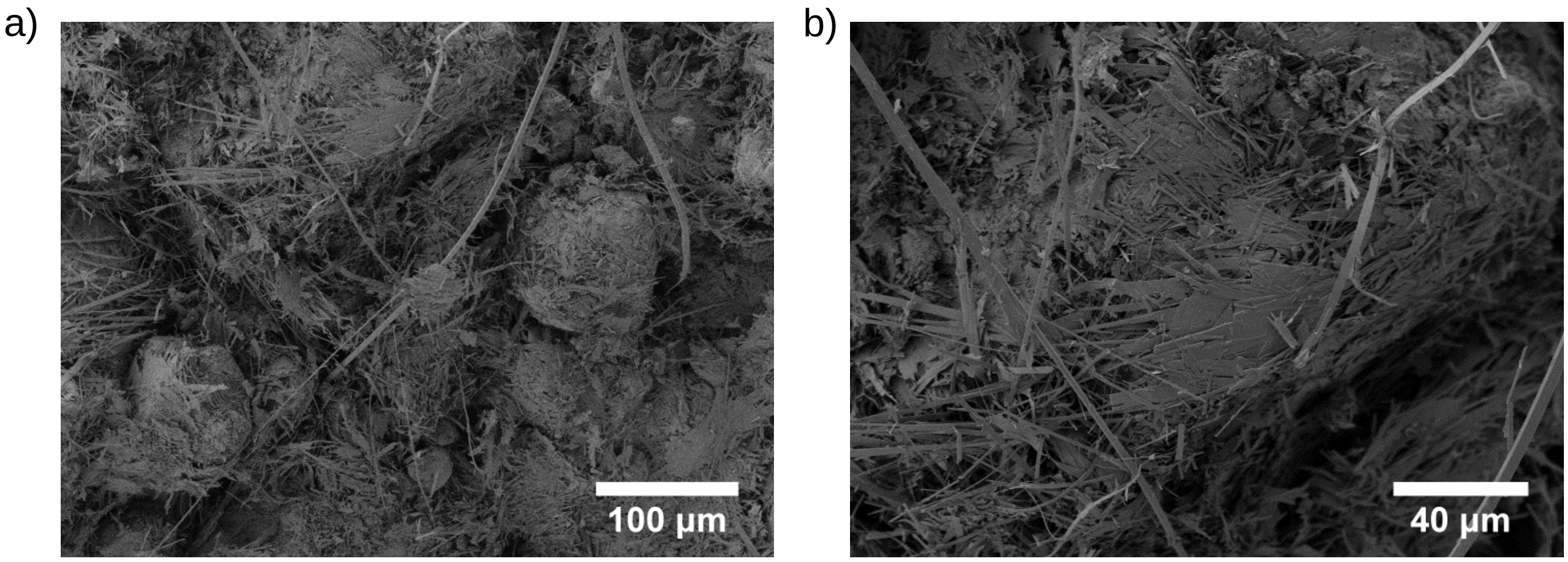

Figure S1: Low-magnification SEM images of bulk $TiS_3$. a) Magnification ×543. b) Magnification ×232.

**RAMAN SPECTROSCOPY**

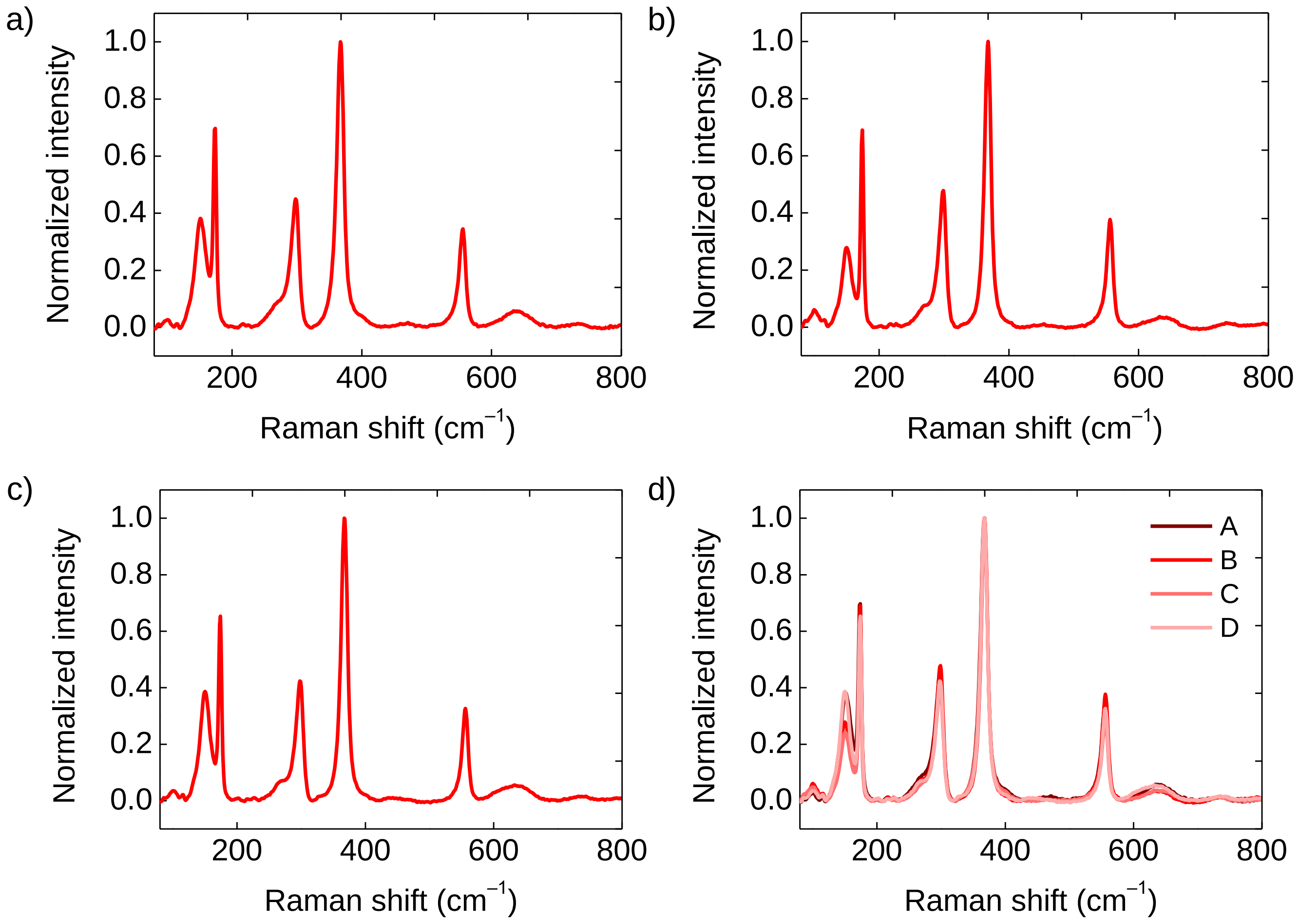

Figure S2: Raman spectra of liquid-phase-exfoliated $TiS_3$ (average of 10 measurements and normalized at 368 cm$^{-1}$). a) Suspension A. b) Suspension B. c) Suspension C. d) Raman spectra of all colloidal suspensions normalized at 368 cm$^{-1}$.

$TiS_3$ bands at 100 cm$^{-1}$ (weak, $B_g$), 174 cm$^{-1}$ (strong, $A_g$), 256 cm$^{-1}$ (shoulder, $B_g$), 273 cm$^{-1}$ (shoulder, $A_g$), 298 cm$^{-1}$ (strong, $A_g$), 368 cm$^{-1}$ (strong, $A_g$), 556 cm$^{-1}$ (strong, $A_g$), 641 cm$^{-1}$ (broad, 273 cm$^{-1}$ + 368 cm$^{-1}$).

Extra bands at 151 cm$^{-1}$ (strong, $E_g$), 201 cm$^{-1}$ (weak, $E_g$), 400 cm$^{-1}$ (weak, $B_{1g}$), 515 cm$^{-1}$ (weak, $B_{1g}$ and $A_{1g}$) and at 641 cm$^{-1}$ (weak, superimposed to $TiS_3$, $E_g$) indicate the presence of $TiO_2$

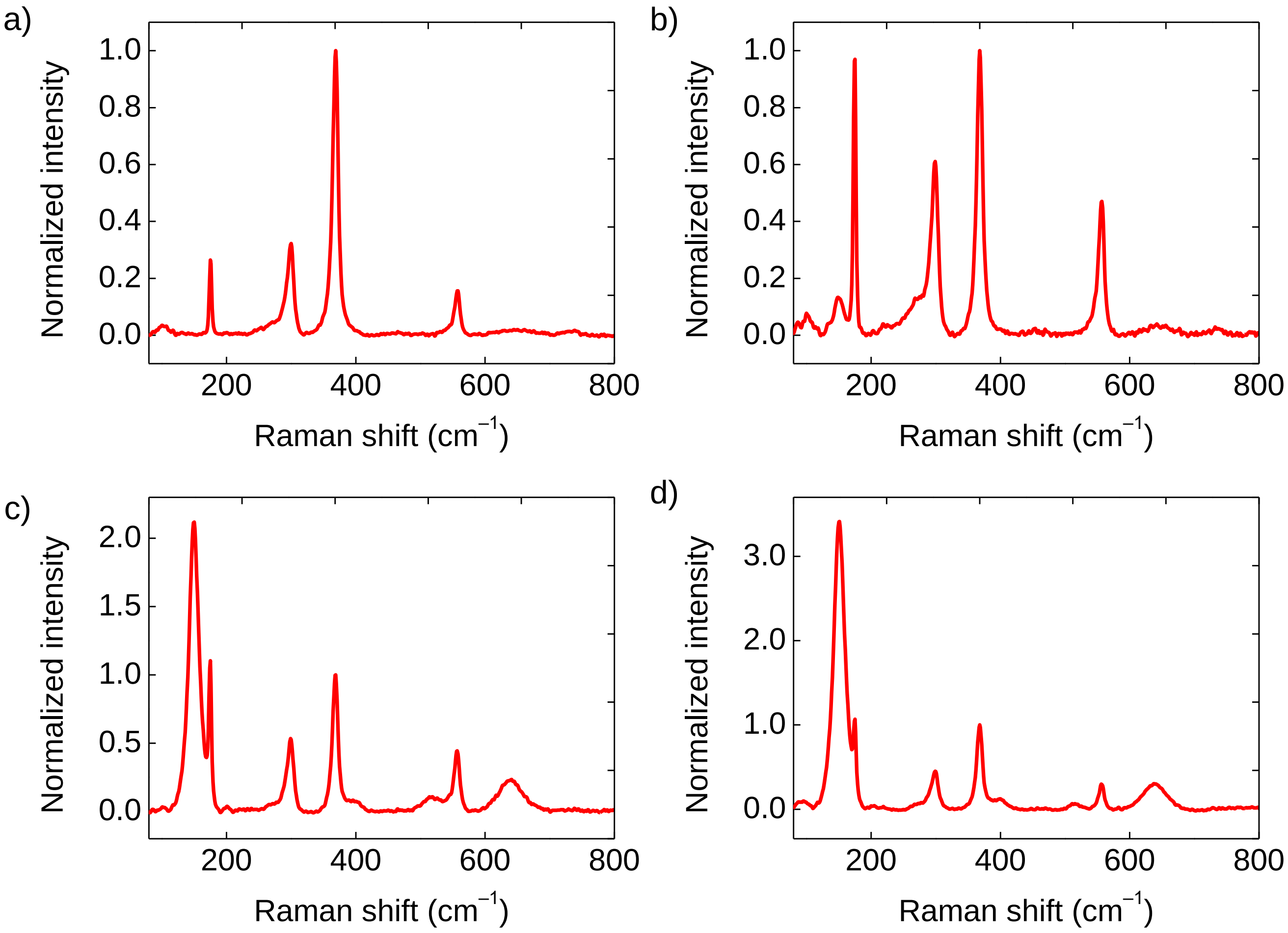

Figure S3: Typical Raman spectra (normalized at 368 cm$^{-1}$) obtained from the analysis of various sites of bulk $TiS_3$.

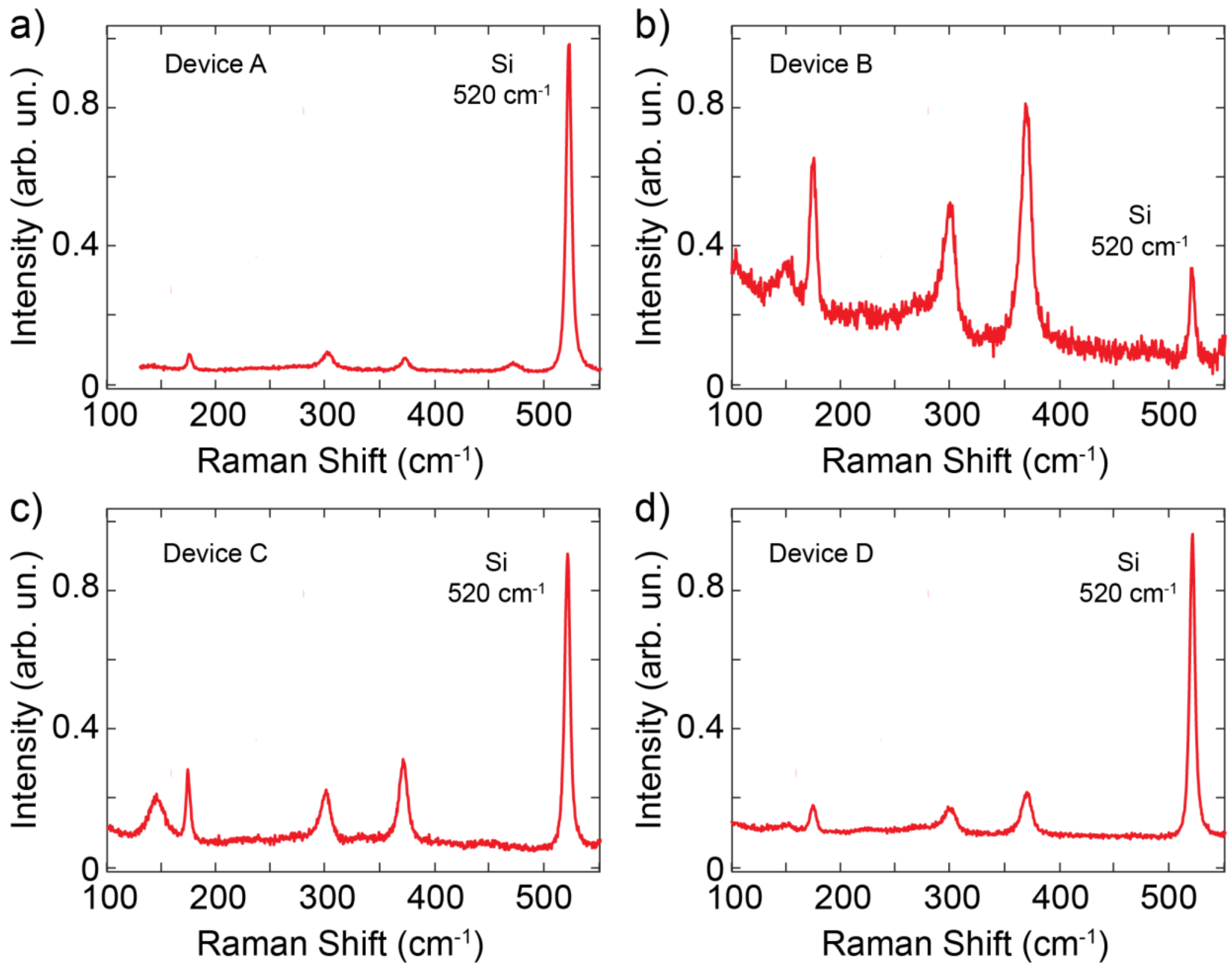

Figure S4: Typical Raman spectra obtained from the analysis of various devices built from liquid-phase exfoliated $TiS_3$. The strong peak at 520 $cm^{-1}$ is a Raman peak of the crystalline silicon substrate.

**UV-Vis-NIR SPECTROSCOPY**

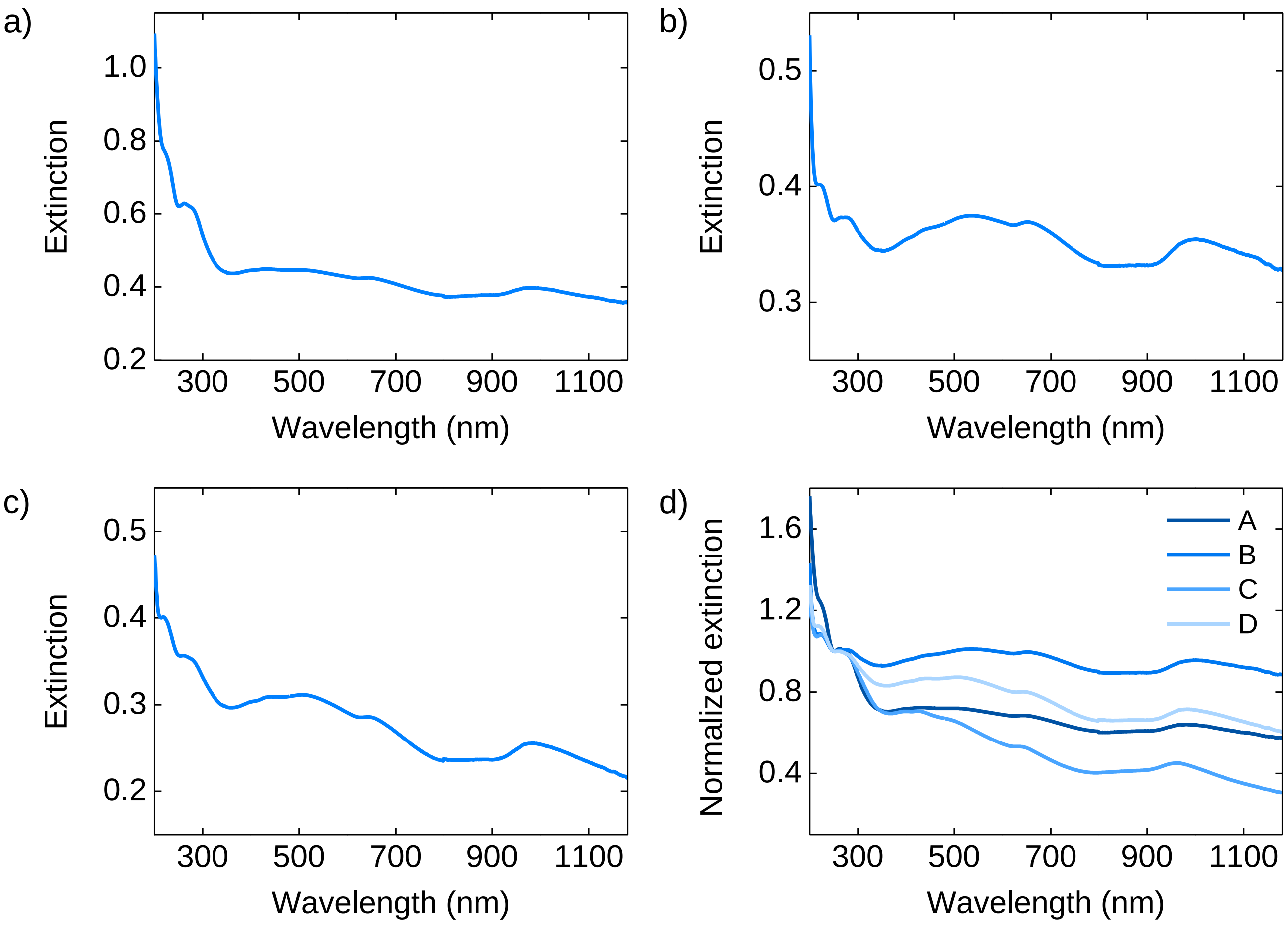

Figure S5: Extinction spectra of liquid-phase-exfoliated $TiS_3$. a) Suspension A. b) Suspension B. c) Suspension C. d) Extinction spectra of all suspensions normalized at 250 nm.

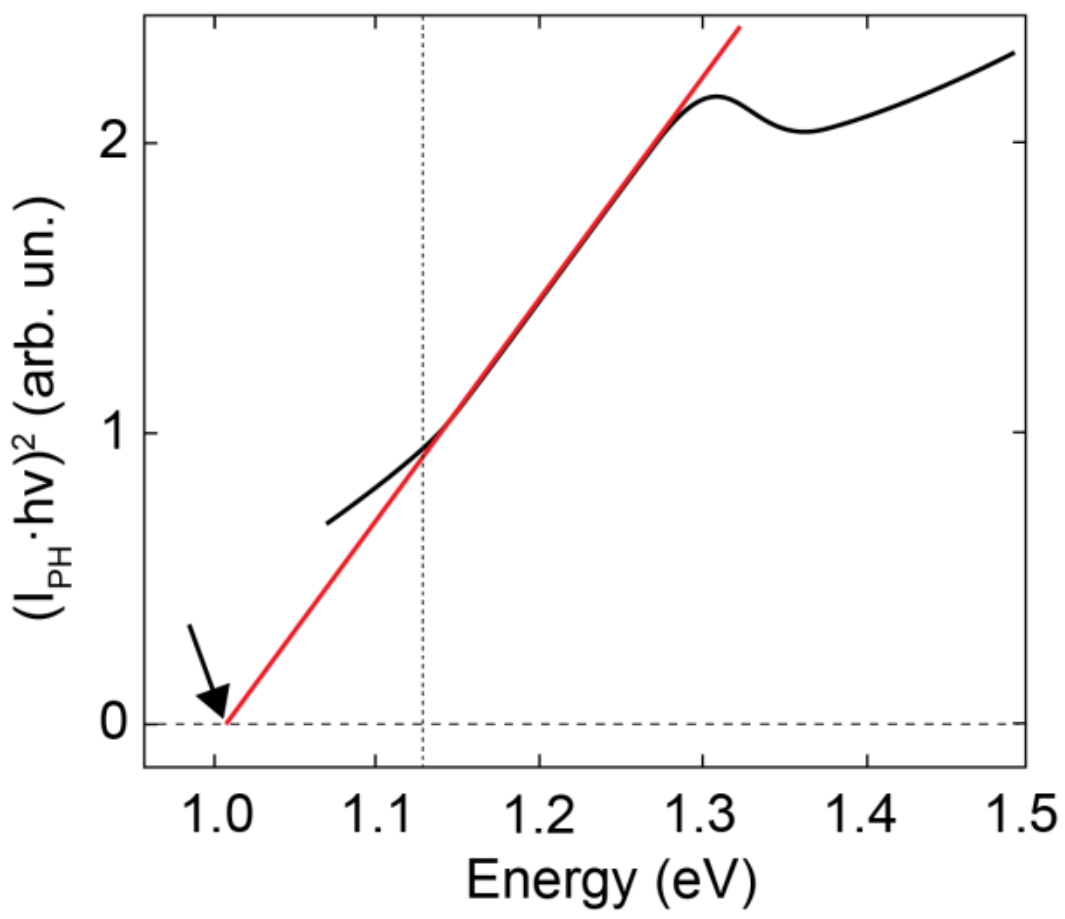

Figure S6. Tauc plot drawn from the absorption data of suspension D. The optical bandgap energy corresponds to the point where the linear fit (red line) meets the X axis.

The absorption studies were limited to wavelengths below 1180 nm due to the strong absorption of IPA that prevents any measurement in the far infrared. Additionally, as the nanoribbons have micrometric length, non-negligible scattering is observed and the resulting spectra must be considered as extinction spectra (that is, the sum of absorption and scattering spectra).

## DIELECTROPHORESIS-BASED LIQUID-PHASE-EXFOLIATED SOLID STATE DEVICES

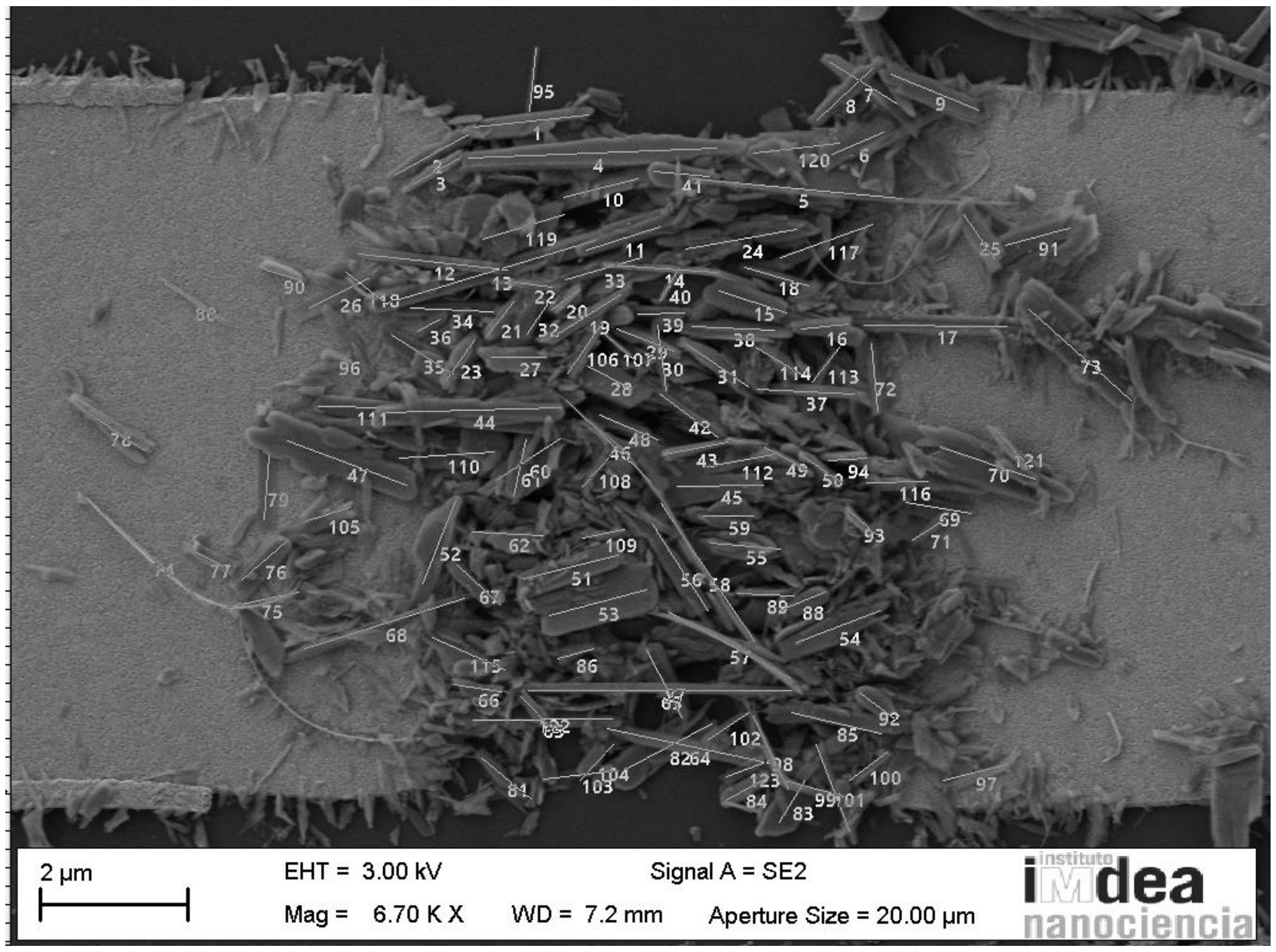

Figure S7: Analysis of the angular distribution of $TiS_3$ nanoribbons between gold electrodes from the device discussed in the main text.

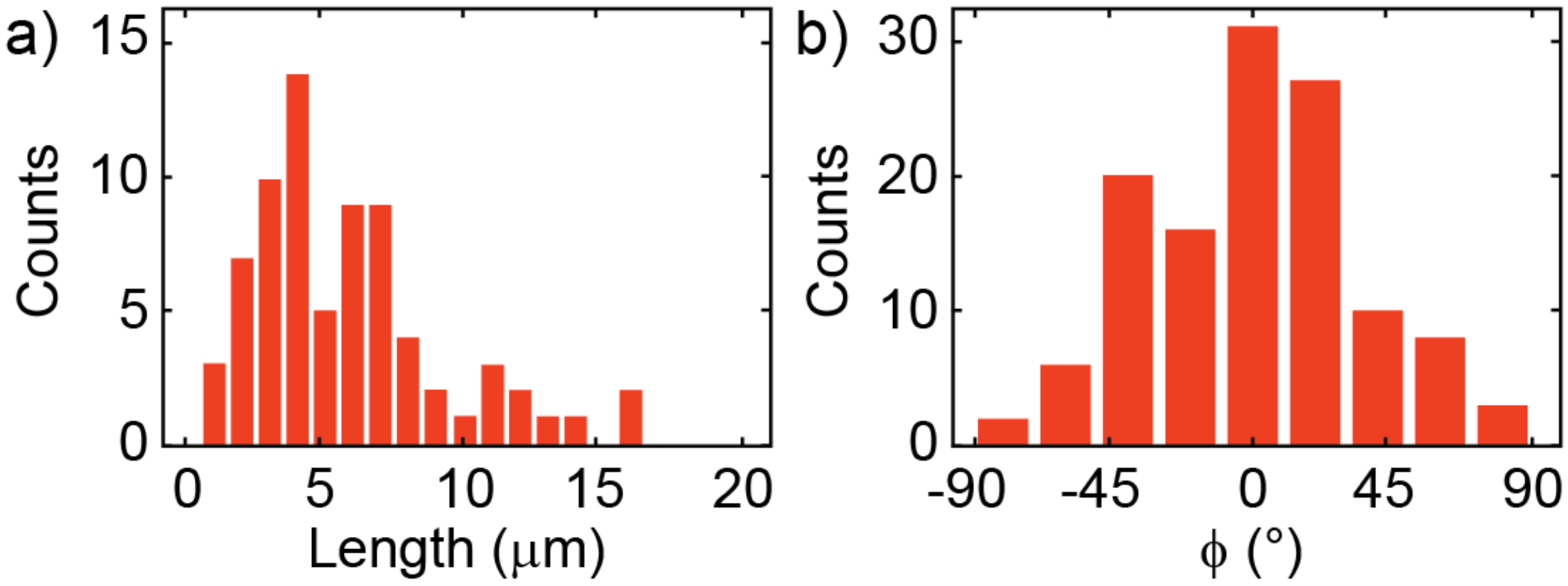

Figure S8: a) Length distribution of the assembled $TiS_3$ nanoribbons. The length was measured from 123 nanoribbons identified in the SEM image of Fig. S6. b) Angular distribution of the assembled $TiS_3$ nanoribbons where 0° correspond to a nanoribbon oriented along the source-drain electrodes direction.

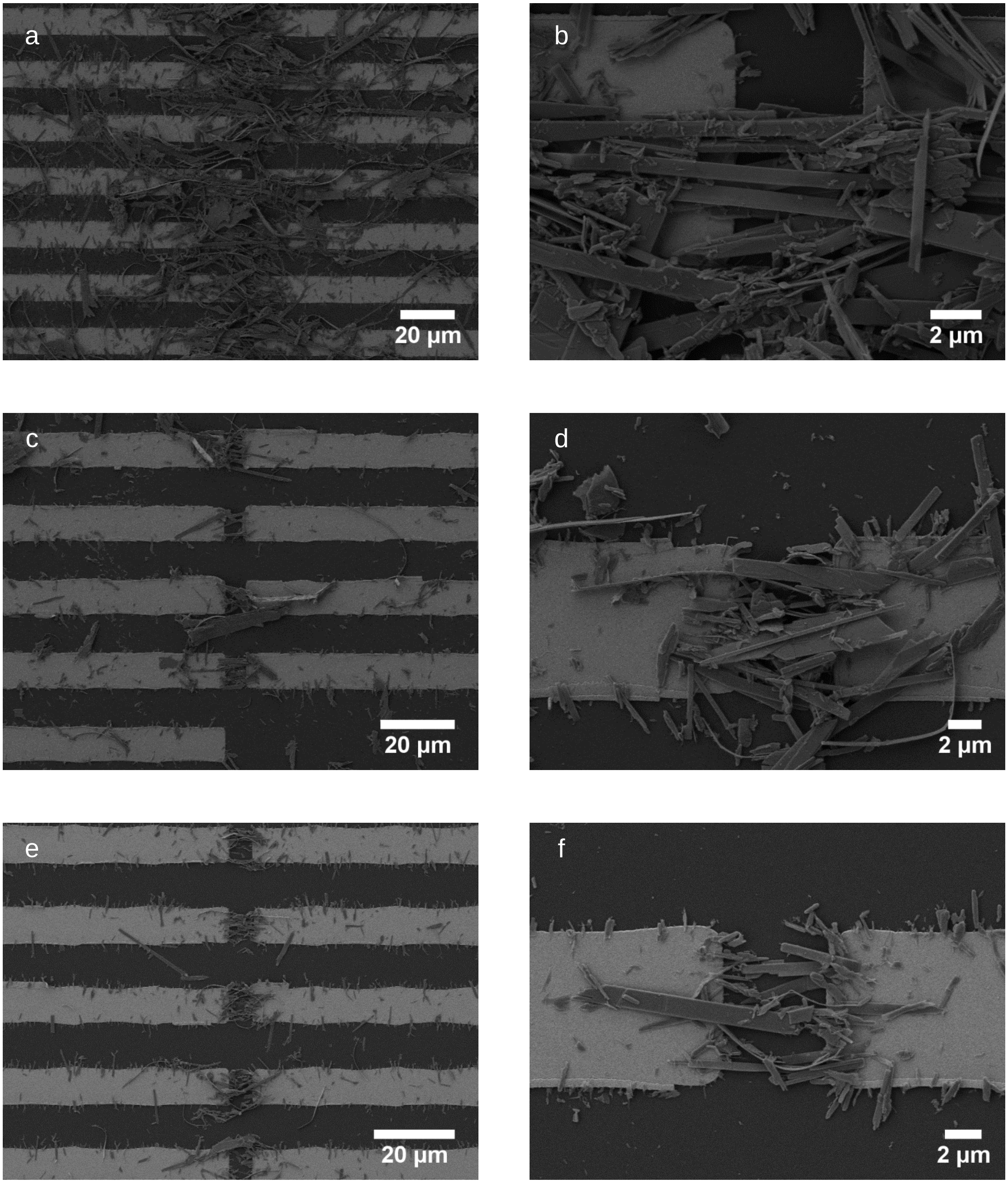

Figure S9: SEM images of the devices made from colloidal suspensions of exfoliated $TiS_3$. a-b) Made from suspension A. c-d) Made from suspension B. e-f) Made from suspension C.

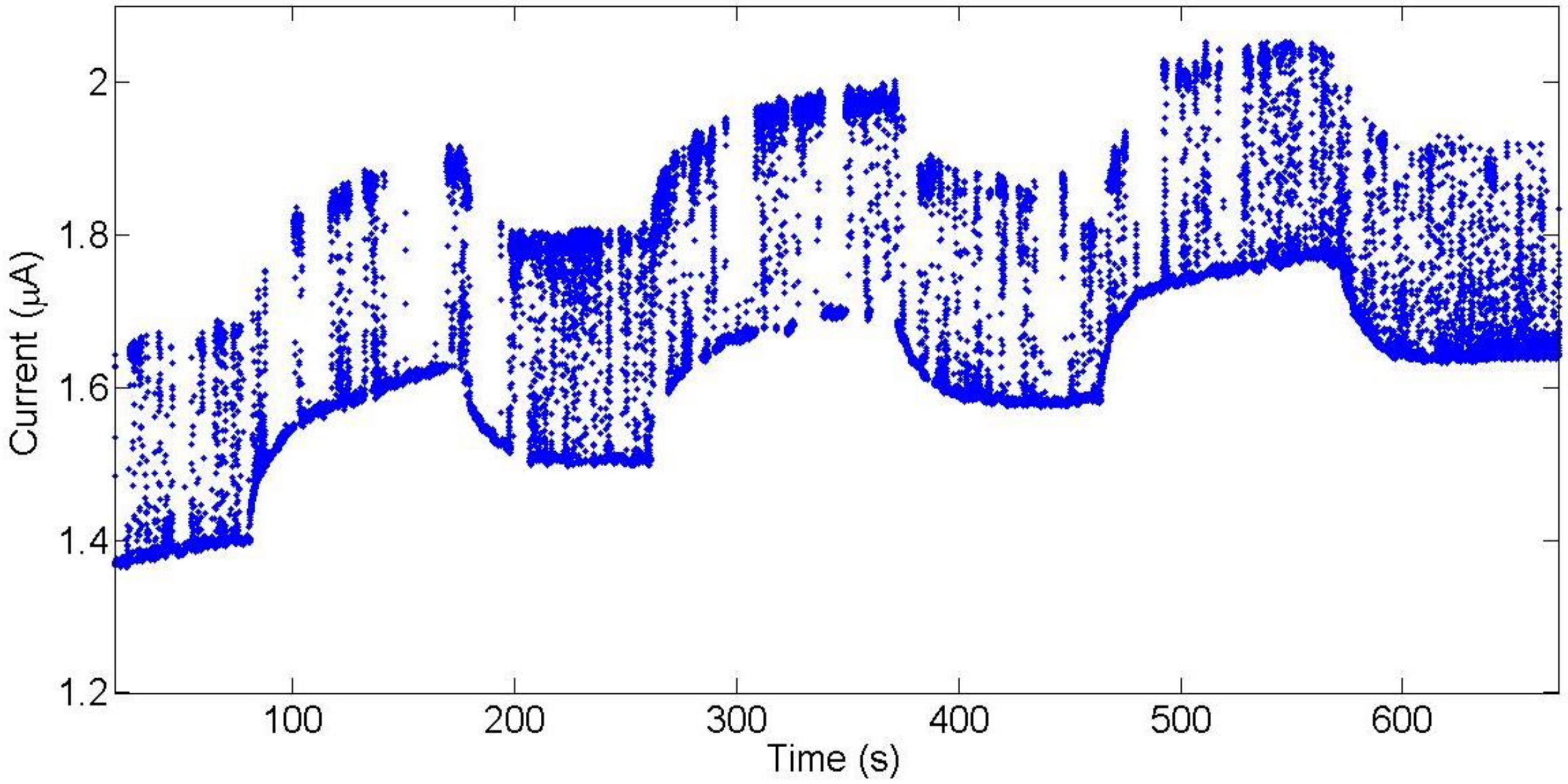

Figure S10: Time response of a device made from suspension A upon modulated optical excitation with a light wavelength of 405 nm.

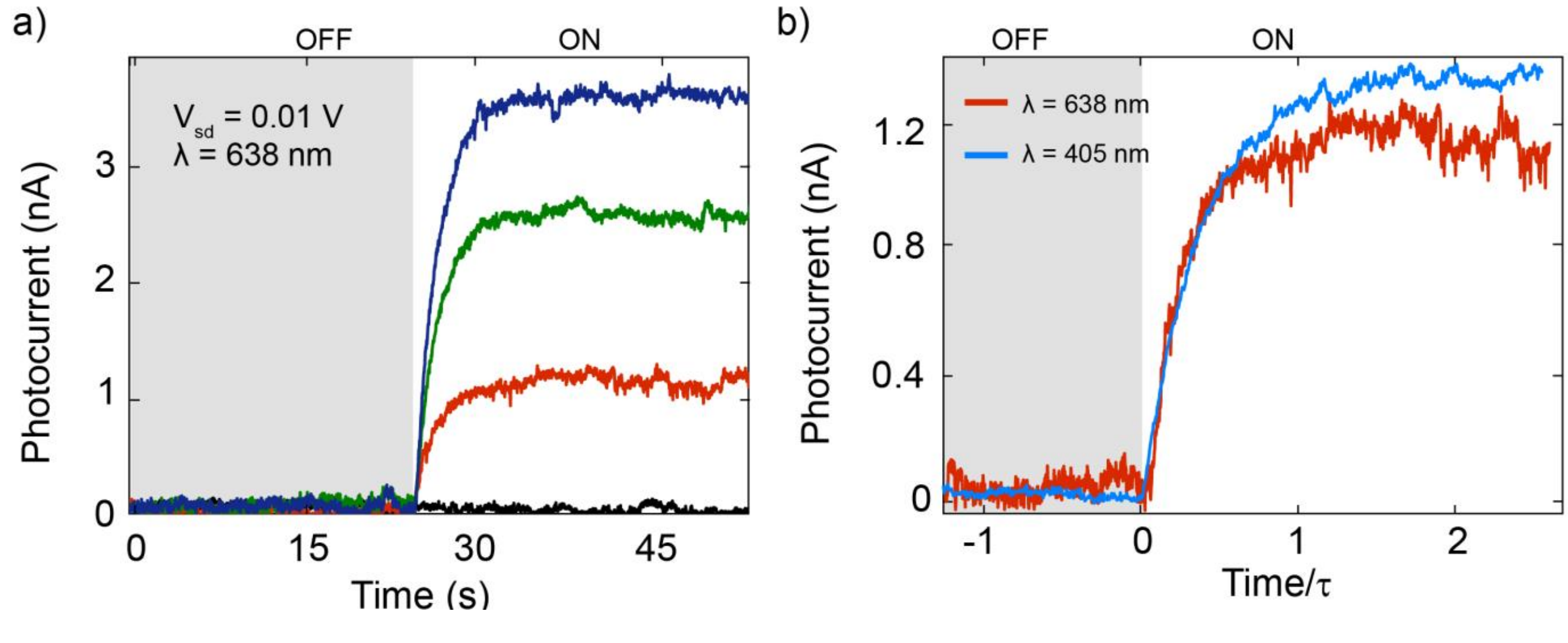

Figure S11: a) Time response of the main text device upon modulated optical excitation with a light wavelength of 638 nm and increasing laser powers up to 47 mW. The measurements are carried out at a bias voltage of 10 mV. b) Time response at two different excitation wavelengths, λ = 405 nm circular spot diameter 120 µm, λ = 638 nm rectangular spot dimensions 100x45 µm. The optical power density of 4 W/mm$^2$ is the same for both illuminations. The time axis of each dataset has been normalized by the time constant (10%-90% rule) of the curve.